\documentclass{PoS}

\usepackage{amssymb}
\usepackage{fontenc}
\usepackage{times}
\usepackage{mathptmx}
\usepackage{graphicx}

\title{Initial conditions for QCD evolution of double parton distributions}

\ShortTitle{Initial conditions for evolution of double parton distributions}

\author{\speaker{Emilia LEWANDOWSKA}%
\\
       Institute of Nuclear Physics PAN\\
       E-mail: \email{emilia.lewandowska@ifj.edu.pl}}

\author{Krzysztof GOLEC-BIERNAT\\
        Institute of Nuclear Physics PAN, University of Rzeszow\\
        E-mail: \email{krzysztof.golec-biernat@ifj.edu.pl}}

\abstract{Double parton distribution functions (DPDFs) are used in the QCD description of double parton scattering. The DPDFs evolve with hard scales through relatively new QCD evolution equations which obey nontrivial momentum and valence quark number sum rules. Based on the constructed numerical program, we present results on the QCD evolution of the DPDFs. In particular, we discuss the problem how to specify initial conditions for the evolution equations which exactly fulfill the  sum rules.}

\FullConference{XXI International Workshop on Deep-Inelastic Scattering and Related Subjects\\
          22-26 April, 2013\\
          Marseilles, France}

\begin{document}

\section{Introduction}

Double parton distribution functions are used in the description of double hard scattering \cite{Blok:2010ge}. Their QCD evolution equations are known in the leading logarithmic approximation (LLA) 
\cite{Snigirev:2003cq,Korotkikh:2004bz,Gaunt:2009re,Ceccopieri:2010kg,Gaunt:2011xd,
Ryskin:2011kk,Diehl:2011tt,Manohar:2012jr}. The DPDFs obey nontrivial sum rules which are conserved by the evolution equations. In this presentation we address the problem  how to specify initial conditions for the evolution equations which exactly obey these sum rules.

\section{Parton distribution functions}
In the single parton scattering (SPS), the final state of the hadron-hadron collision has been produced from only one hard interaction while in the double parton scattering, two hard subprocesses occur. 
For the description of the SPS we use the single parton distribution functions, ${D_f(x,Q)}$, while for the double parton scattering - the double parton distribution functions denoted by ${D_{f_1f_2}(x_1,x_2,Q_1,Q_2)}$. The DPDF depend on  parton flavours ${f_1,f_2}$ (including gluon), longitudinal momentum fractions ${x_1,x_2}$ and two hard scales ${Q_1,Q_2}$.
The  parton momentum fractions obey the condition, 
\begin{eqnarray}
\label{eq:limit}
x_1+x_2 \le 1\,,
\end{eqnarray}
which says that the sum of partons' momenta  cannot exceed the total nucleon momentum

\section{Evolution equations}

The general form of QCD evolution equations for single PDFs is given by 
\begin{eqnarray}
\label{eq:onepdfeq}
\partial_{t}D_{f}(x,t)=\sum_{f^\prime}\int^{1}_{0}du\, {\cal{K}}_{ff^\prime}(x,u,t)\,D_{f^\prime}(u,t),
\end{eqnarray}
with the evolution parameter ${t=\ln(Q^2/Q_0^2})$. 
The integral kernels ${{\cal{K}}_{ff^\prime}}$ presented in above equation describe real and virtual parton emission. 
The real emission kernels take the following form
\begin{eqnarray}
{\cal{K}}^R_{ff'}(x,u,t)=\frac{1}{u} P_{ff'}(\frac{x}{u},t)\,\theta(u-x)
\end{eqnarray}
in which ${P_{ff'}}$ are  splitting functions computed perturbatively in QCD in powers of the strong coulpling constant ${\alpha_s}$:
\begin{eqnarray}
P_{ff'}(z,t)=\frac{\alpha_s(t)}{2\pi}P^{(0)}_{ff'}(z)+\frac{\alpha^2_s(t)}{(2\pi)^2}P^{(1)}_{ff'}(z)+...
\end{eqnarray}
After including the splitting functions, we find the well known DGLAP evolution equations for the single PDFs:
\begin{eqnarray}\label{eq:dglap}
\partial_{t}D_{f}(x,t)=\sum_{f^\prime}\int^{1}_{x}\frac{dz}{z}P_{ff'}(z,t)\,D_{f'}(\frac{x}{z},t)-D_f(x,t)\sum_{f'}\int^1_0 dz\,z\, P_{f'f}(z,t).
\end{eqnarray}

In the leading logarithmic approximation,
the evolution equations of the DPDFs  (for equal two hard scales, $Q_1=Q_2,\equiv Q$) have the following form, see \cite{Gaunt:2009re} for more details,
\begin{eqnarray}\label{eq:ddglap}
\nonumber
\label{eq:twopdfeq}
\partial_t\, D_{f_1f_2}(x_1,x_2,t) &=& \sum_{f'}\int^{1-x_2}_{0} du \,{\cal{K}}_{f_1f'}(x_1,u,t) \, D_{f' f_2}(u,x_2,t)
\\\nonumber
&+& \sum_{f'}\int_{0}^{1-x_1}du\,{\cal{K}}_{f_2f'}(x_2,u,t) \,D_{f_1f'}(x_1,u,t)
\\
&+& \sum_{f'}\,{\cal{K}}_{f'\to f_1f_2}^R (x_1,x_1+x_2,t)\, D_{f'}(x_1+x_2,t).
\end{eqnarray}
The upper integration limits reflect condition (\ref{eq:limit}).
The third term constains the single PDFs and because of that eqs. (\ref{eq:twopdfeq}) and (\ref{eq:onepdfeq}) have to be solved together. That is why the initial conditions for both the single  and double PDFs have to be specified at 
some initial scale $Q_0$.

\section{Sum rules}

The DGLAP evolution equations (\ref{eq:dglap}) preserve
the momentum sum rule for the single PDFs:
\begin{eqnarray}
\sum_f \int^1_0 dx\,x\,D_f(x,Q)=1\,
\end{eqnarray}
while the evolution equations (\ref{eq:ddglap}) preserve
the momentum sum rule for the DPDFs
\begin{eqnarray}
\sum_{f_1}\int_{0}^{1-x_2}dx_1\,x_1\,  \frac{D_{f_1f_2}(x_1,x_2,Q)}{D_{f_2}(x_2,Q)}=(1-x_2).
\end{eqnarray}
The ratio of the double and single PDFs in the above relation looks like a conditional probability to find a parton with the momentum fraction ${x_1}$, while the second parton is fixed. Thus, it is clearly seen that the new momentum sum rule relates the  double and single PDFs
\begin{eqnarray}
\label{eq:momrule}
\sum_{f_1}\int_{0}^{1-x_2}dx_1\,x_1\,D_{f_1f_2}(x_1,x_2,Q)=(1-x_2)\,D_{f_2}(x_2,Q).
\end{eqnarray}

The valence quark  number sum rule for the single PDFs has the well known form
\begin{eqnarray}
\int_0^{1}dx\left\{D_{q_i}(x,Q)-D_{\bar{q_i}}(x,Q)\right\}=N_{i}
\end{eqnarray}
where $N_i$ is the number of valence quarks.
For the DPDFs, the following relation holds, depending   on the second parton flavour, \cite{Gaunt:2009re}
\begin{eqnarray}
\int_0^{1-x_2}dx_1\left\{D_{q_if_2}(x_1,x_2,Q)-D_{\bar{q_i}f_2}(x_1,x_2,Q)\right\}
=\left\{
\begin{array}{ll}
   N_{i}\,D_{f_2}(x_2,Q)      &      \mbox{\rm ~~~~for $f_2\ne q_i,\bar{q_i}$} \\  \\
(N_{i}-1)\,D_{f_2}(x_2,Q)      &      \mbox{\rm ~~~~for $f_2=q_i$} \\ \\
(N_{i}+1)\,D_{f_2}(x_2,Q)      &     \mbox{\rm ~~~~for $f_2=\bar{q_i}$}\,
\end{array}
\right.
\label{eq:valrule}
\end{eqnarray}
It is important  to emphasize again that the momentum and valence quark number sum rules are conserved by the evolution equations (\ref{eq:dglap}) and (\ref{eq:ddglap}) once they are imposed at an initial scale $Q_0$.

\section{Initial conditions}

In order to solve eqs.~(\ref{eq:dglap}) and (\ref{eq:ddglap}) we need to specify initial conditions for the DPDFs. For practical reason, their form 
is to built  using the existing single PDFs. For example, in \cite{Korotkikh:2004bz, Gaunt:2009re} a {\it symmetric}  input with respect to the parton interchange
was proposed,
\begin{eqnarray}\label{eq:gsinput}
D_{f_1f_2}(x_1,x_2,Q_0)=D_{f_1}(x_1,Q_0)\,D_{f_2}(x_2,Q_0)\,\frac{(1-x_1-x_2)^2}{(1-x_1)^{2+n_1}(1-x_2)^{2+n_2}}\,,
\end{eqnarray}
which is also positive definite (provided the single PDFs are  positive). 
In the below, we show how the momentum and valence quark number sum rules are
fulfilled by this input by  showing the ratios of the r.h.s to l.h.s of 
eqs.~(\ref{eq:momrule}) and (\ref{eq:valrule}) for $q_i=u$ (and $n_1=n_2=0$ in (\ref{eq:gsinput}), for simplicity). 
We see that the valence quark number   sum rule is significantly violated.
\begin{center}
\vskip -0.6cm
\includegraphics[height = 6cm]{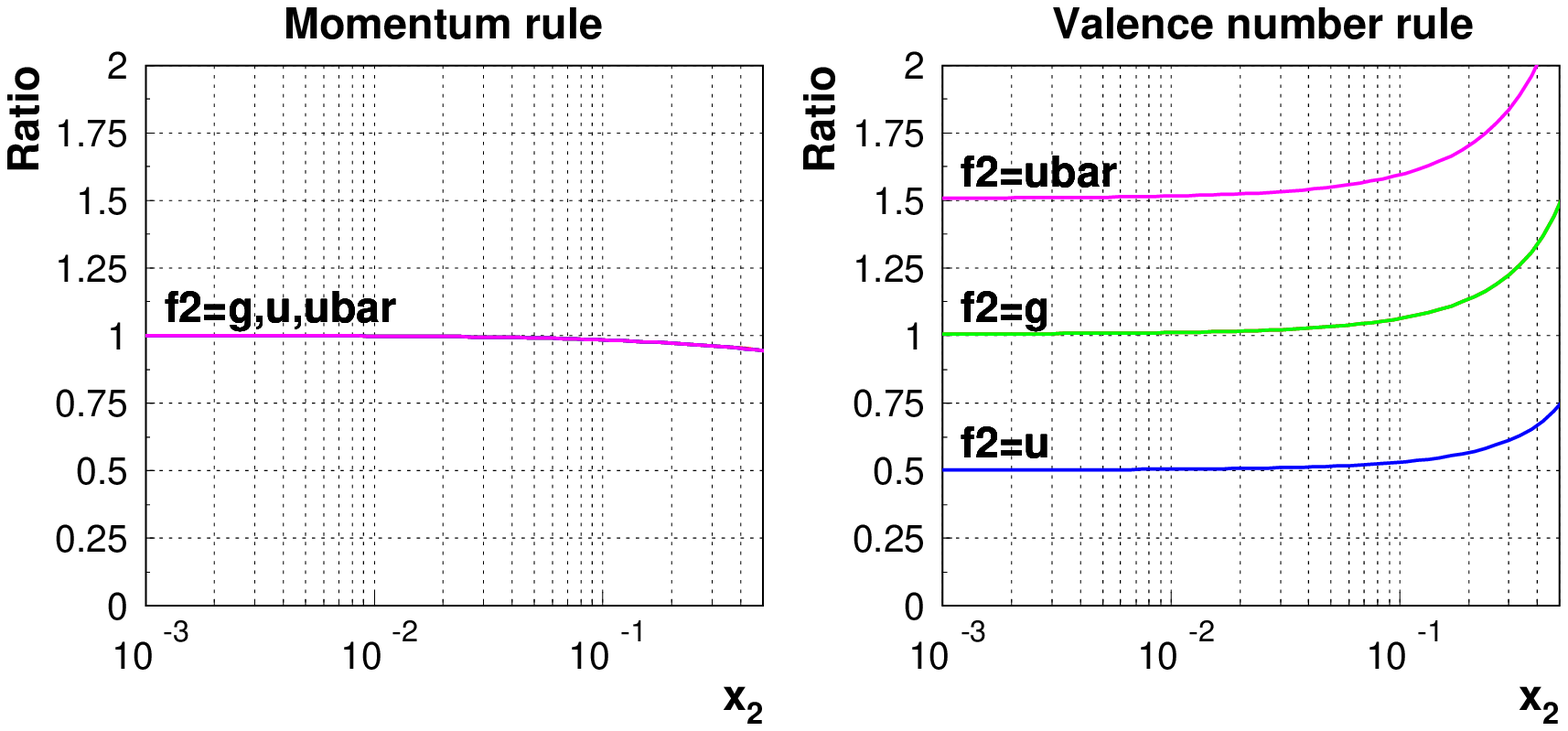}
\end{center}

Is it possible to construct an input 
which {\it exacly} fulfill the sum rules (\ref{eq:momrule}) and (\ref{eq:valrule})?
In order to obey the momentum sum rule we could use an {\it asymmetric} ansatz (we skip 
$Q_0$ in the notation): 
\begin{eqnarray}
D_{f_1f_2}(x_1,x_2)\,=\, \frac{1}{1-x_2} D_{f_1}\!\left(\frac{x_1}{1-x_2}\right)\cdot D_{f_2}(x_2).
\end{eqnarray}
To fulfill the valence number sum rule we need to introduce corrections for identical quark flavours and antiflavours,
 \begin{eqnarray}\nonumber
D_{f_if_i}(x_1,x_2)\!\!\! &=& \!\!\! \frac{1}{1-x_2} \left\{D_{f_1}(\frac{x_1}{1-x_2}) - \frac{1}{2}
\right\} D_{f_1}(x_2) 
\\\nonumber
 \\
D_{f_i\bar{f_i}}(x_1,x_2)\!\!\! &=&  \!\!\!\frac{1}{1-x_2} \left\{D_{f_1}(\frac{x_1}{1-x_2}) + \frac{1}{2}
\right\} D_{\bar{f_1}}(x_2)\,,
\end{eqnarray}
which  do not spoil the already fulfilled momentum sum rule. However, we pay
the price that the DPDFs for identical flavours are not positive definite  because of the factor $-1/2$. We cannot avoid such a situation in the contruction which uses
single PDFs.  
\newpage
The graphical comparison of the symmetric and asymmetric inputs is shown below.
\begin{center}
\includegraphics[height = 8.5cm]{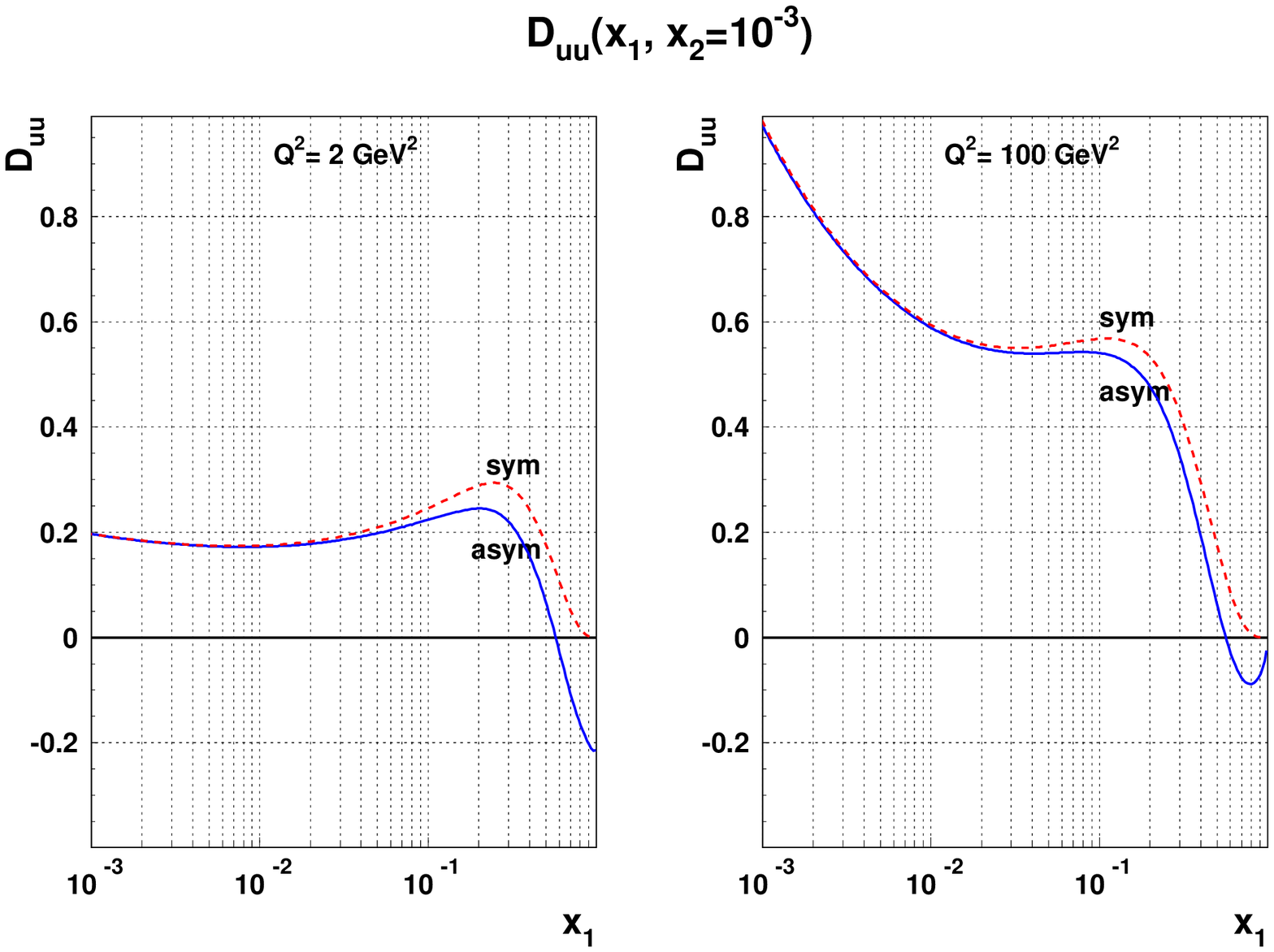}
\end{center}
\vskip -0.4cm
For the distributions ${D_{uu}}$ both inputs give similar results in the small ${x_1}$ region, while for  large ${x_1}$ there are differences between them because of the lack of  positive definitness of the asymmetric input. The same results are found for the  evolved distribution. 

For ${D_{u\bar{u}}}$, both distributions are positive but there are differences between them at the large ${x_1}$: 
\begin{center}
\includegraphics[height = 8.5cm]{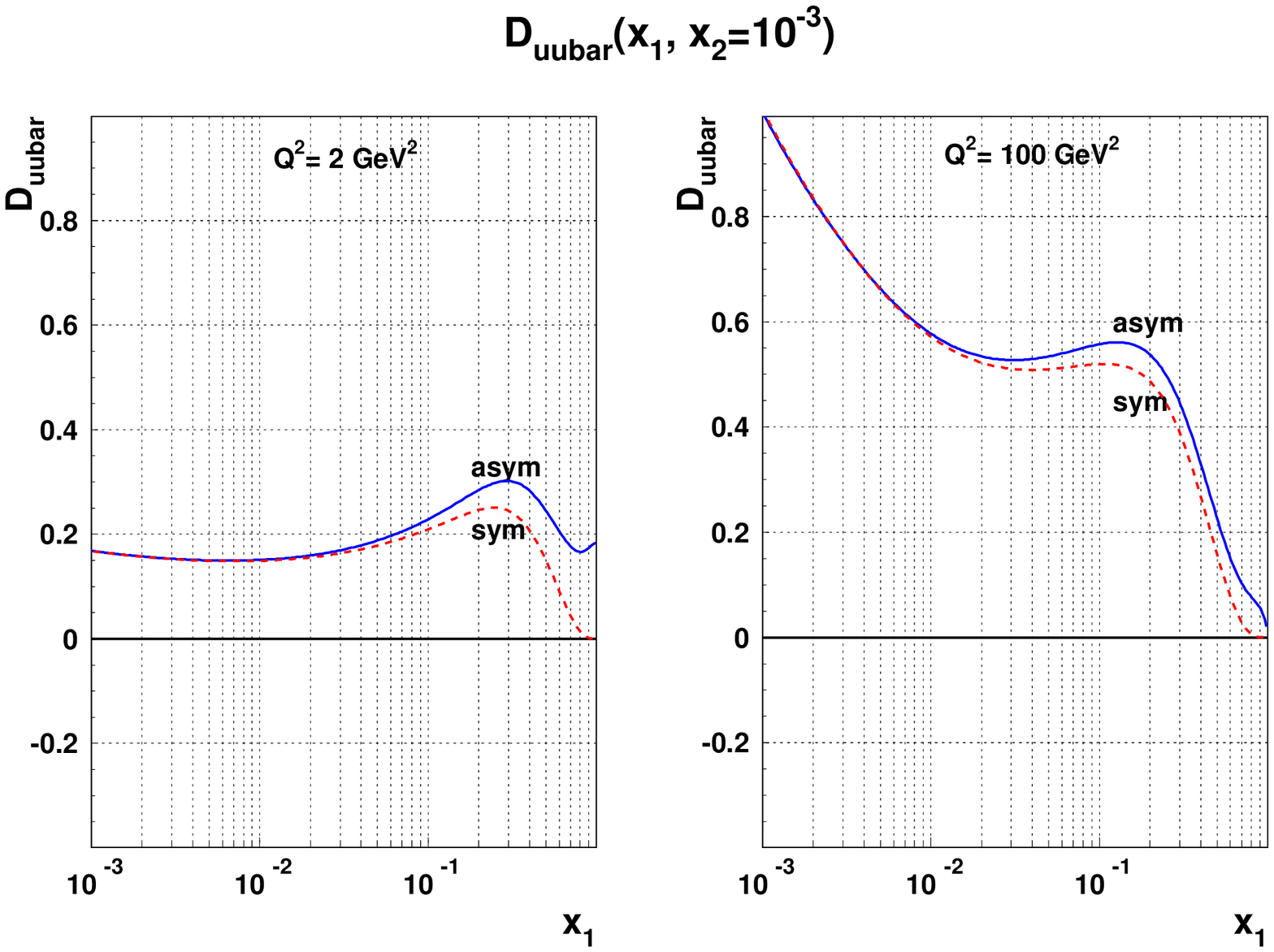}
\end{center}
\newpage
And finally, for the ${D_{gu}}$ distributions both inputs give the same results in the whole ${x_1}$ domain. 
\begin{center}
\includegraphics[height = 8.5cm]{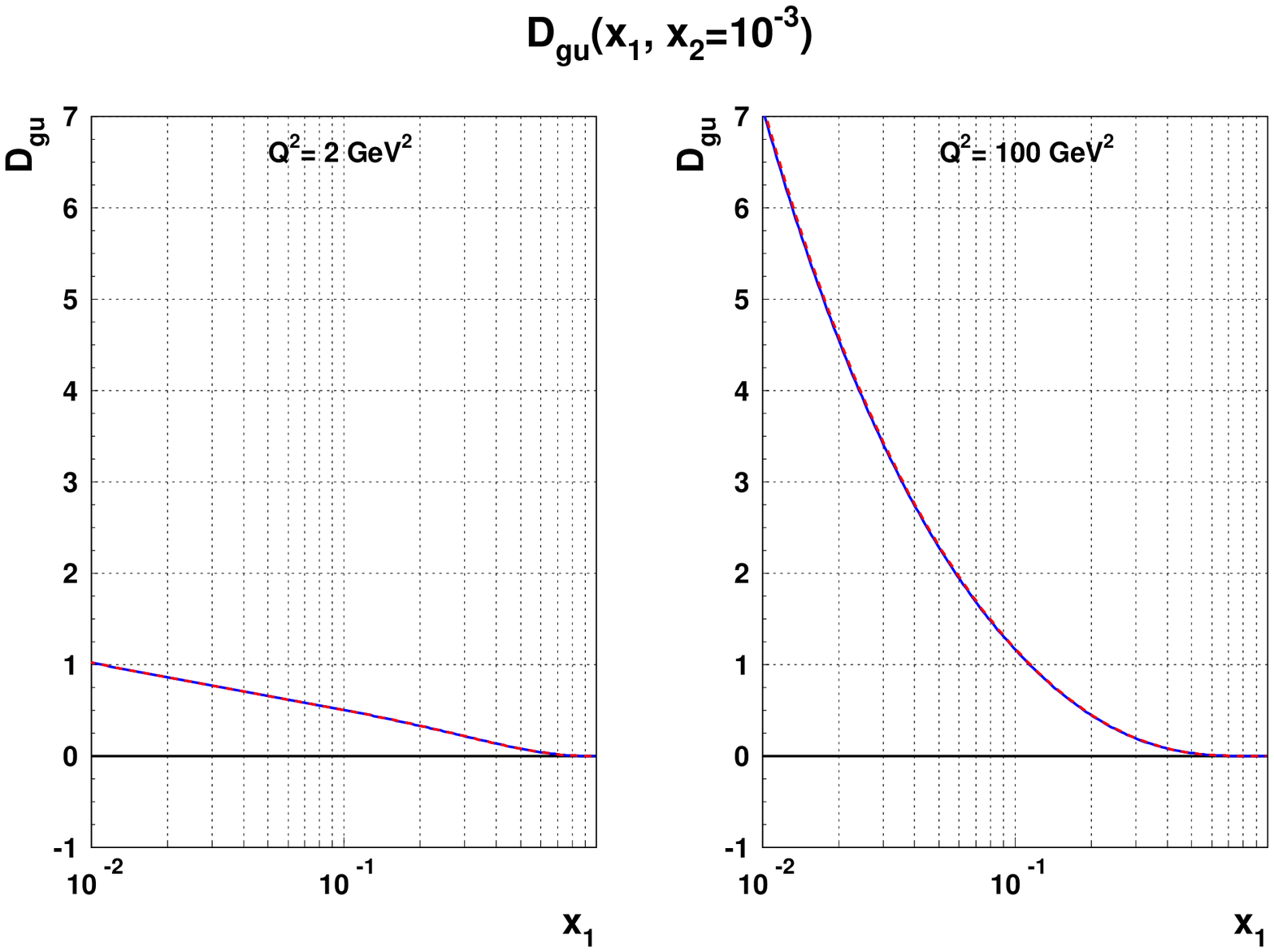}
\end{center}

\section{Summary}

The specification of the initial conditions for evolution of the DPDFs is not a simple task. The symmetric input obeys parton symmetry and positivity but does not exactly fulfill sum rules. On the other hand, the asymmetric input obeys the sum rules exactly but it is not positive definite.  There also exist an alternative solution: one could specify positive initial double distributions and then generate the single PDFs using the sum rules. However, a little is known  about the DPDFs from experiments. 
For small values of parton momentum fractions, the factorized form, ${D_{f_1f_2}(x_1,x_2,Q)\approx D_{f_1}(x_1,Q)\,D_{f_2}(x_2,Q)}$, is a good approximation. However, the problem  occurs for large values of $x$ ($>10^{-2})$.

\medskip
\centerline{\bf Acknowledgement}
This work was supported by the Polish NCN grants DEC-2011/01/B/ST2/03915  and  DEC-2012/05/N/ST2/02678.


\bibliographystyle{h-physrev4}
\bibliography{mybib}

\end{document}